\documentclass[sigconf]{acmart}

\def\BibTeX{{\rm B\kern-.05em{\sc i\kern-.025em b}\kern-.08emT\kern-.1667em\lower.7ex\hbox{E}\kern-.125emX}}

\usepackage{tikz}
\usepackage{amsmath}
\usepackage{algorithm}
\usepackage{algorithmic}
\usepackage{balance}
\usepackage{hyperref}
\usepackage[margin=0.2cm]{caption}
\usepackage{multirow}
\usepackage[symbol]{footmisc}
\usepackage{bbding}
\usepackage{array}
\usepackage{pifont}
\usepackage{url}
\usepackage{subfigure}
\usepackage{colortbl}
\usepackage[T1]{fontenc}
\usepackage{aecompl}

\newcommand{\oursys}{{EasyScale}}
\newcommand{\ourth}{{EasyScaleThread}}
\newcommand{\ourcompany}{{CompanyA}}

\newcommand{\eg}{{e.g.}}
\newcommand{\ie}{{i.e.}}
\newcommand{\etc}{{etc.}}
\newcommand{\aka}{{a.k.a.}}

\copyrightyear{2023}
\acmYear{2023}
\setcopyright{acmlicensed}\acmConference[SC '23]{The International Conference for High Performance Computing, Networking, Storage and Analysis}{November 12--17, 2023}{Denver, CO, USA}
\acmBooktitle{The International Conference for High Performance Computing, Networking, Storage and Analysis (SC '23), November 12--17, 2023, Denver, CO, USA}
\acmPrice{15.00}
\acmDOI{10.1145/3581784.3607054}
\acmISBN{979-8-4007-0109-2/23/11}

\begin{document}

\title{EasyScale: Elastic Training with Consistent Accuracy and Improved Utilization on GPUs}

\author{Mingzhen Li}
\affiliation{\institution{Beihang University} \city{Beijing} \country{China}}
\email{lmzhhh@buaa.edu.cn}

\author{Wencong Xiao}
\affiliation{\institution{Unaffiliated} \city{Hangzhou} \country{China}}
\email{xiaowencong@gmail.com}

\author{Hailong Yang}
\affiliation{\institution{Beihang University} \city{Beijing} \country{China}}
\email{hailong.yang@buaa.edu.cn}

\author{Biao Sun}
\affiliation{\institution{Beihang University} \city{Beijing} \country{China}}
\email{biaosun@buaa.edu.cn}

\author{Hanyu Zhao}
\affiliation{\institution{Unaffiliated} \city{Hangzhou} \country{China}}
\email{zhaohanyu1994@gmail.com}

\author{Shiru Ren}
\affiliation{\institution{Unaffiliated} \city{Beijing} \country{China}}
\email{renshiru2000@gmail.com}

\author{Zhongzhi Luan}
\affiliation{\institution{Beihang University} \city{Beijing} \country{China}}
\email{07680@buaa.edu.cn}

\author{Xianyan Jia}
\affiliation{\institution{Unaffiliated} \city{Hangzhou} \country{China}}
\email{jiaxianyan@gmail.com}

\author{Yi Liu}
\affiliation{\institution{Beihang University} \city{Beijing} \country{China}}
\email{yi.liu@buaa.edu.cn}

\author{Yong Li}
\affiliation{\institution{Unaffiliated} \city{Beijing} \country{China}}
\email{relianceslee@gmail.com}

\author{Wei Lin}
\affiliation{\institution{Unaffiliated} \city{Hangzhou} \country{China}}
\email{ustcwlin@hotmail.com}

\author{Depei Qian}
\affiliation{\institution{Beihang University} \city{Beijing} \country{China}}
\email{depeiq@buaa.edu.cn}


%
\renewcommand{\shortauthors}{Li and Xiao, et al.}

%

%

\begin{abstract}

Distributed synchronized GPU training is commonly used for deep learning.
The resource constraint of using a fixed number of GPUs makes large-scale training jobs suffer from long queuing time for resource allocation, and lowers the cluster utilization.
Adapting to resource elasticity can alleviate this but often introduces inconsistent model accuracy, due to lacking of capability to decouple model training procedure from resource allocation. 
We propose EasyScale, an elastic training system that achieves consistent model accuracy under resource elasticity for both homogeneous and heterogeneous GPUs.
EasyScale preserves the data-parallel training behaviors strictly, traces the consistency-relevant factors carefully, utilizes the deep learning characteristics for EasyScaleThread abstraction and fast context-switching. 
To utilize heterogeneous cluster, EasyScale dynamically assigns workers based on the intra-/inter-job schedulers, minimizing load imbalance and maximizing aggregated job throughput. 
Deployed in an online serving cluster, EasyScale powers the training jobs to utilize idle GPUs opportunistically, improving overall cluster utilization by 62.1\%.

\end{abstract}
%
%


%

%

%
\maketitle


\section{Introduction} 
\label{sec:introduction}

Deep learning (DL) is now playing a vital role in supporting a wide range of indispensable applications,
such as advertising for online shopping, computer vision for autonomous driving, 
natural language processing for searching, \etc{}
Recognizing the promising power of DL, large companies have built large-scale shared GPU clusters to expedite the adoption of DL in almost every production scenario. 
The common practice today often adopts distributed deep learning training (DLT),
where each worker typically processes training data in mini-batches and uses synchronized stochastic gradient descent (Sync-SGD) to compute gradients for model update. 
For example, PyTorch~\cite{paszke2019pytorch} typically allocates a GPU per worker and uses Distributed Data Parallel (DDP) to perform gradient synchronization across mini-batches.
However, a DLT job will not start until all resources become available simultaneously due to the gang-scheduling~\cite{antman2020,jeon2019analysis}.
Besides, the DLT job is executed in a fixed degree of parallelism (DoP), 
thus can never scale in/out when fluctuating GPU resources become available due to cluster load change.
The fixed DoP prevents the DLT jobs from adapting to the resource elasticity that is common in shared GPU clusters~\cite{borg, weng2022mlaas}.

Recently, a series of researches (\eg{}, TorchElastic~\cite{torchelastic}, ElasticDL~\cite{elasticdl}, VirtualFlow~\cite{virtualflow2021}, Pollux~\cite{pollux2021}) have proposed elastic training frameworks that allow a DLT job to continue its training procedure under resource elasticity. 
Several cluster management approaches (\eg{}, Gandiva~\cite{gandiva2018}, Optimus~\cite{optimus}, KungFu~\cite{kungfu2020}) have also utilized the resource elasticity to maximize cluster utilization or allocate job resources for fast training convergence.
Despite the well-known benefit of elasticity, the proposed elastic training frameworks have rarely been used in the industry so far. 
Because they changes the hyper-parameters and the training procedure according to available resources, and thus introduces the non-determinism during training and inevitably affects model accuracy, \eg, overall accuracy and per-class accuracy. 
The fundamental obstacle for their adoption is the \textit{inconsistent model accuracy} when training with different resources (\S\ref{sec:motivation}), which is problematic and may destroy the model usability under elastic training, so that it prevents DL practitioners from embracing elastic training.

When training DL models, the DL developers usually go through two separate stages. \textit{1)} Model designing: the model architecture together with hyper-parameters (\eg{}, learning-rate / batch-size / optimizer) are determined by developers. \textit{2)} Model training: the model is executed to fit the training dataset over epochs. 
Given the same output of model designing stage (\aka, a certain training job), the training results can be reproduced through model training with the same number of fixed GPUs.
However, prior works~\cite{torchelastic,elasticdl,virtualflow2021,pollux2021} require DL developers to partially delegate model designing stage to the elastic training frameworks (\eg, allow Pollux to tune learning-rate adaptively and allow TorchElastic to tune batch-size proportionally), which changes the hyper-parameters and the training procedure explicitly or implicitly.

To overcome the issues mentioned above, we propose EasyScale (\S\ref{sec:design}), the first elastic training system that achieves consistent model accuracy under resource elasticity for both homogeneous and heterogeneous GPU resources, 
thereby improving the overall cluster efficiency by utilizing the idle GPUs at best effort for elastic model training.
Compared to other works, EasyScale has two distinct features: \textit{1)} it faithfully preserves all DL developer's intentions in model designing stage, while benefiting DL training jobs with resource elasticity in model training stage; \textit{2)} it avoids introducing extra non-determinism from resource elasticity (for number of GPUs and heterogeneity in GPU types). 
The goal of EasyScale is to erase the non-determinism in the model training stage of DL training and ensure the accuracy of elastic training consistent with fixed DoP training.

During DL model training, EasyScale \textit{treats rigorous determinism and reproducibility as the first-class goal}. 
\oursys{} explores the possibilities of producing \textit{bitwise-consistent} model regardless of the number and type of GPU resources allocated. 
The fundamental idea of \oursys{} is to decouple the distributed model training procedure from hardware resource allocation.
This is done by an abstraction called \textit{\ourth{}} (EST in short, \S\ref{sec:easyscalethread}), which encapsulates all the training stages such as data loading, sampling, computation, and communication. The EST abstraction enables the training behaviors under resource elasticity are exactly the same as executed under fixed resources. 
To minimize the abstraction overhead, \oursys{} utilizes the DL characteristics for fast context switching across ESTs, efficient on-demand checkpointing of EST states when the resource scales, and optimized data loading worker sharing along the EST execution.
To eliminate the non-determinism in resource elasticity and heterogeneity, \oursys{} sources the root causes scattered across the software stack of DL training, and then controls them through embedding the implicit states in EST contexts / checkpoints and fixing others (\S\ref{sec:nondeterminism}).
In addition, \oursys{} introduces both \textit{intra-job} and \textit{inter-job} schedulers regarding the EST abstraction, to improve the utilization of GPU resources and the aggregated throughput of the entire cluster (\S\ref{sec:policy}).

\oursys{} is implemented by modifying a
popular framework, PyTorch, to provide the elastic training capability without compromising model accuracy (\S\ref{sec:impl}). In addition, the scheduling policies of \oursys{} are implemented on top of Kubernetes scheduler.
We evaluate \oursys{} on a cluster with 64 heterogeneous GPUs to demonstrate its effectiveness in accuracy-consistent model training using micro-benchmarks on typical workloads (\S\ref{sec:benchmark_exp}). We also show the advantage of \oursys{} under resource elasticity with production workload trace (\S\ref{sec:trace_exp}).
The trace experiment shows that \oursys{} can generate consistent model accuracy for all DLT jobs compared to those using a specific number of GPUs. 
In addition, \oursys{} improves the average job completion time (JCT) by 13.2$\times$ and makespan by 2.8$\times$ thanks to its intra-job and inter-job schedulers.
We have deployed \oursys{} in a production cluster equipped with 3,000+ heterogeneous GPUs for online model serving (\S\ref{subsec:cluster_exp}).
The evaluation result demonstrates that \oursys{} improves 
the GPU allocation ratio by 17.1\% and the average GPU utilization by 62.1\%. 
The evaluation result also shows that using \oursys{}, the DLT jobs can automatically scale in seconds when co-located with online model serving jobs.

Specifically, the key contributions are as follows:
\begin{itemize}
\item 
We propose the \oursys{} framework for elastic distributed model training to achieve consistent model accuracy.
It utilizes EST abstraction to preserve consistent training behaviors as PyTorch DDP, and achieves efficient context-switching under resource elasticity.
\item 
We investigate the non-deterministic behaviors of model training in existing elastic training frameworks,
and identify factors scattered across the entire DLT software stack that affect the bitwise accuracy of model training.
\item 
We propose the \oursys{} scheduler, including intra-job and inter-job schedulers to improve the utilization of heterogeneous GPU resources of the entire cluster, regarding the EST abstraction.
\item 
We deploy \oursys{} in production clusters to co-locate elastic training jobs with online model serving jobs. The evaluation results show that \oursys{} significantly improves cluster utilization. 
\end{itemize}


\section{Motivation}
\label{sec:motivation}

In this section, we first briefly describe the increasing demand for adopting elastic resources when training DL models on large-scale shared GPU clusters.
We then analyze the non-determinism of existing elastic training frameworks to motivate \oursys{}.

\begin{figure*}[t]
    \begin{minipage}[t]{0.21\textwidth}
        \centering
        \includegraphics[width=\linewidth]{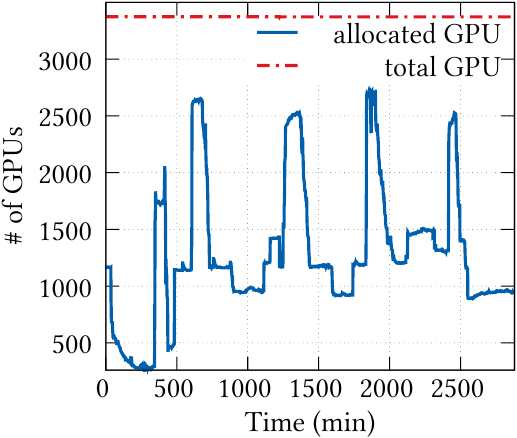} 
        \caption{Online serving GPU cluster load variation.}
        \label{fig:onlineinferenceload} 
    \end{minipage}
    \begin{minipage}[t]{0.24\textwidth}
        \centering 
        \includegraphics[width=0.95\linewidth]{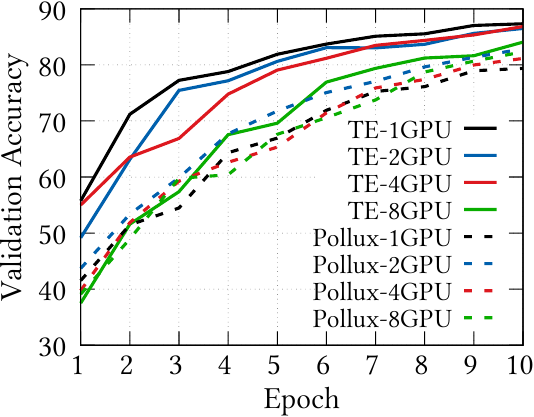} 
        \caption{Non-deterministic accuracy curves of ResNet18.} 
        \label{fig:elasticaccuracy} 
    \end{minipage}%
    \begin{minipage}{0.30\textwidth}
        \vspace{-0.6in}
        \centering
        \fontsize{6}{6}\selectfont
        \setlength\tabcolsep{0.8pt}
        \setlength\extrarowheight{1.9pt}
        \begin{tabular}{ccccccccccccc}
            \hline
            &      & C 0  & C 1  & C 2  & C 3  & C 4  & C 5  & C 6  & C 7  & C 8  & C 9  & Total \\ \hline
            \multirow{5}{*}{\rotatebox{90}{\ \ \ TorchElastic}} & 1GPU & 92.5 & 96.7 & 87.1 & 85.2 & 93.7 & 85.7 & 95.8 & 95.3 & 95.3 & 93.9 & 92.1 \\
            & 2GPU & 89.1 & 96.2 & 91.3 & 82.5 & 94.4 & 82.9 & 95.3 & 93.4 & 93   & 96.7 & 91.5 \\
            & 4GPU & 89   & 93.4 & 89.3 & 84.7 & 94.7 & 86.8 & 94.7 & 92.7 & 95.6 & 97.4 & 91.8 \\ 
            & 8GPU & 90.8 & 95.5 & 85.7 & 81.6 & 92.6 & 90.3 & 93.3 & 97.1 & 95   & 95.3 & 91.7 \\  \cline{2-13} 
            \rowcolor{gray!50} & \textbf{$\Delta$} & 3.5  & 3.3  & 5.6  & 3.6  & 2.1  & \textbf{7.4}  & 2.5  & 4.4  & 2.6  & 3.5  & \textbf{0.6}  \\ \hline
            \multirow{5}{*}{\rotatebox{90}{\ \ \ Pollux}}       & 1GPU & 93.9 & 96.7 & 87.6 & 86.1 & 95   & 85.7 & 92.5 & 94.8 & 96.7 & 93.9 & 92.3 \\ 
            & 2GPU & 90.2 & 95.9 & 87   & 77.8 & 95.5 & 80.1 & 94.4 & 96.9 & 96.2 & 93.2 & 90.7 \\ 
            & 4GPU & 92.6 & 97.6 & 75.6 & 81.7 & 84.8 & 89.7 & 94.9 & 91.4 & 96.7 & 92.3 & 89.7 \\ 
            & 8GPU & 91.7 & 98.3 & 92.9 & 84   & 79   & 82.7 & 91.7 & 93   & 96.1 & 85   & 89.4 \\ \cline{2-13} 
            \rowcolor{gray!50} & \textbf{$\Delta$} & 3.7  & 2.4  & \textbf{17.3} & 8.3  & 16.5 & 9.6  & 3.2  & 5.5  & 0.6  & 8.9  & \textbf{2.8}  \\ \hline
        \end{tabular}
        \caption{Non-deterministic per-class accuracy of ResNet18 on CIFAR10 at epoch 100.}
        \label{tab:per-class-accuracy}
    \end{minipage}%
    \begin{minipage}[t]{0.24\textwidth} 
        \centering 
        \includegraphics[width=0.9\linewidth]{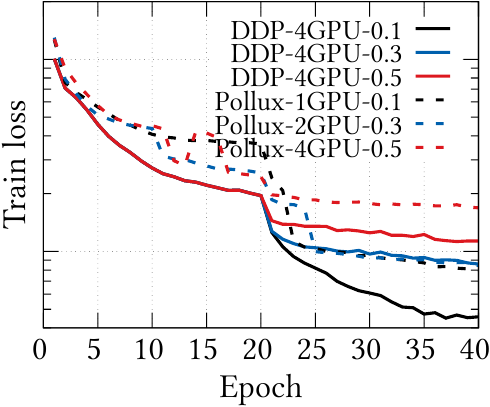} 
        \caption{Train loss of ResNet50 with different hyper-parameter gamma.} 
        \label{fig:ddpvspollux} 
    \end{minipage} 
\end{figure*}

\subsection{Demand for Adapting to Resource Elasticity}
\label{sec:motivation-elasticity}

Distributed training is widely adopted in production clusters for processing massive data and large models.
However, it usually achieves sub-optimal performance on shared clusters for the following reasons. For one reason, large-scale DLT jobs commonly suffer from long queuing time for resource allocation due to gang-scheduling policy~\cite{jeon2019analysis, jeon2018multi, hived2020, antman2020, weng2022mlaas}. For the other reason, the training can be easily interrupted due to constant resource revocation by high-priority jobs. 
Our 2-day statistic in a GPU production cluster of \ourcompany{} shows that
jobs requesting more than 8 GPUs account for 61.7\% of training failures due to resource revocation.
Whereas the jobs requesting one GPU only account for 5.3\% of training failures due to resource revocation. This discrepancy can be attributed to the inherent Sync-SGD training behavior, where terminating any worker stops the entire model training.

To avoid frequent failure when training DL models on large-scale shared cluster, it is important to establish the ability to adapt the training procedure to resource elasticity (\aka, elastic training). Another benefit of elastic training is that a multi-GPU job can start the training immediately with the required number of GPUs gradually allocated, and thus eliminate the long queuing time due to gang scheduling.
Moreover, the elastic training also reveals more opportunities to utilize the idle resources of the online model serving cluster for training DLT jobs.
Our 2-day GPU allocation statistics of an online model serving cluster (Figure~\ref{fig:onlineinferenceload}) show that the difference of required number of GPUs between idle and peak hours can be up to 2,000.
Ideally, the idle GPUs can be shared by both model training and online serving to improve GPU utilization, 
similar to the big-data workloads~\cite{borg}. 
Exploiting such an opportunity also demands elastic training to meet the SLAs of online serving jobs.

\subsection{Non-determinism over Elastic Training}
\label{sec:motivation-nondeterminism}
To address the drawbacks described above,
a series of research works~\cite{elasticdl,torchelastic,virtualflow2021, pollux2021} 
have proposed elastic training frameworks that enable adapting to dynamically scaling resources at runtime. 
So that training jobs can start executing with available resources as soon as possible, thus eliminating the mandatory queuing time and avoiding the frequent failure, which improves cluster utilization and reduces job completion time.
Existing elastic training frameworks usually adopt optimized synchronization methods, such as gradient accumulation~\cite{virtualflow2021, pollux2021}), hyper-parameter tuning~\cite{kungfu2020}, and batch size adjustment~\cite{torchelastic}), to eventually reach similar model accuracy compared to the training with static resources.
However, they still have rarely been used in industry due to the following drawbacks.

\textbf{Inconsistent Model Accuracy --} The multiple runs of model training with elastic training frameworks fail to generate consistent model accuracy when using different amounts of resources.
Figure~\ref{fig:elasticaccuracy} illustrates the validation accuracy of training ResNet18 model on CIFAR10 dataset, with varying numbers of V100 GPUs.
We keep all hyper-parameters and random seeds as default except for using different allocated GPUs. 
TorchElastic (TE)~\cite{torchelastic} is configured with linear scaling rule for adjusting learning rates~\cite{imagenet1hour},
and Pollux~\cite{pollux2021} can automatically decide the learning rate and batch size accordingly.
It is clear that resource elasticity leads to different training behaviors compared to model training on a fixed number of GPUs.
The result also shows that Pollux introduces less accuracy variance compared to TorchElastic, 
however the difference is still non-negligible (\eg, up to 5.8\% at epoch 10).

To better understand the inconsistent model accuracy with elastic training, we extend the training using TorchElastic and Pollux to 100 epochs, and report the overall and per-class (10 classes in total) accuracy in Figure~\ref{tab:per-class-accuracy}. 
The overall accuracy variance for TorchElastic and Pollux is still notable, with 0.6\% and 2.8\%, respectively. Note that even the latest elastic training framework VirtualFlow~\cite{virtualflow2021} also suffers from 0.4\% accuracy degradation on ResNet50 according to its experiments. However, we cannot provide a direct comparison with VirtualFlow since its implementation is not publicly available. The per-class accuracy variance for TorchElastic and Pollux is even larger, reaching up to 7.4\% and 17.3\% maximally, whereas 3.9\% and 7.4\% on average.
The per-class accuracy variance can be detrimental to the model usability, 
in scenarios such as life-critical pedestrian detection for self-driving cars~\cite{gerasimou2020importance} and profit-critical recommendation/advertising systems~\cite{zhang2019deep}. 
Moreover, the upper bound of the variance/inconsistency for model accuracy remains unknown, which further increases the hesitation of DL practitioners in adopting elastic training.

\textbf{Difficult to Understand Hyper-parameter Effect --} 
Model developers conduct model training to seek better hyper-parameters and model structure, and reason their effectiveness through deterministic reproducibility on fixed GPUs.
Figure~\ref{fig:ddpvspollux} shows the experiment of ResNet50 on CIFAR10 by comparing the elastic training on 1/2/4 GPUs using Pollux~\cite{pollux2021} to the non-elastic training on fixed 4 GPUs using PyTorch DDP.
The configurations remain the same except the hyper-parameter of learning rate (gamma), 
which decides the learning rate decay factor after certain training epochs (\eg, 20 epochs in this experiment).
The PyTorch DDP is conducted on 4 GPUs with the gamma value set to 0.1, 0.3, and 0.5 for each run. 
The Pollux is conducted on 1/2/4 GPUs with gamma of 0.1/0.3/0.5, respectively. 
As illustrated in Figure~\ref{fig:ddpvspollux}, when using PyTorch DDP, it is clear to identify the trend of how the hyper-parameter gamma affects the training loss during training. 
However, when training with different numbers of GPUs, the loss curves of Pollux have many unexpected oscillations and thus reveal no clear trend for DL developers, which invalidates the existing knowledge of hyper-parameter tuning, and thus hinders the productivity of model designing stage. 

To summarize, we believe that the non-determinism over elastic training leads to inconsistent model accuracy and complicates the hyper-parameter tuning. The fundamental reason can be attributed to that, the existing elastic training frameworks~\cite{elasticdl, horovodelastic, torchelastic, kungfu2020, pollux2021,virtualflow2021} lack the capability to decouple the resource allocation from the model training procedure (\S\ref{sec:nondeterminism}), and thus fail to provide consistent model accuracy during resource elasticity. 
In order to effectively utilize the elastic resources on shared GPU clusters without compromising the model accuracy, we argue that a new elastic training framework should be proposed to address the non-determinism across the DLT software stack for achieving consistent model accuracy. The new elastic training framework should also provide further opportunities to improve the throughput of individual DLT jobs as well as the utilization of the entire GPU cluster.


\section{Design}
\label{sec:design}

\subsection{Overview}

\begin{figure*}[t]
    \begin{minipage}[t]{0.27\linewidth}
        \centering
        \includegraphics[width=0.95\linewidth]{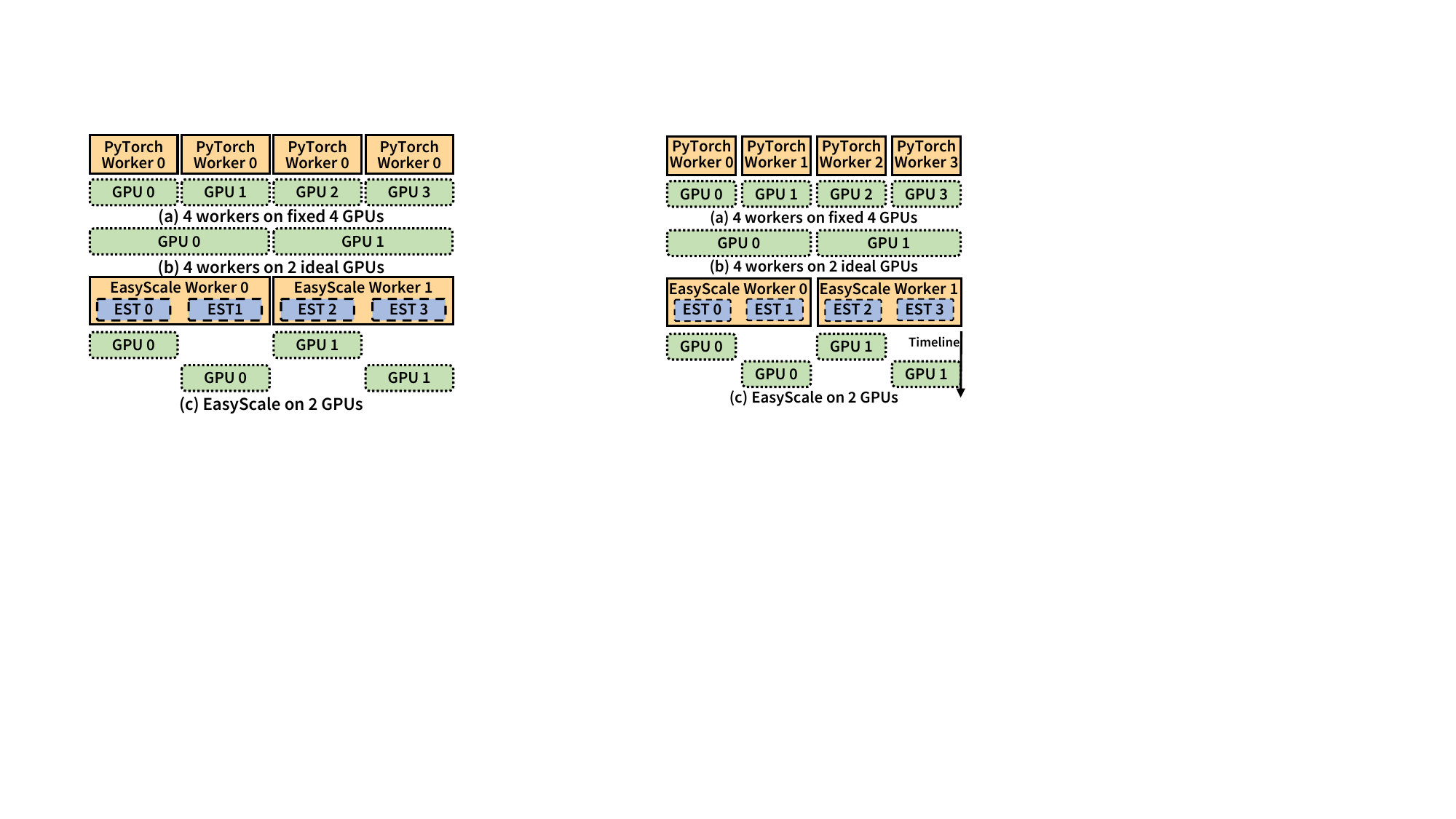}
        \caption{Idea of EasyScale.}
        \label{fig:overview}
    \end{minipage}
    \begin{minipage}[t]{0.72\linewidth}
        \centering
        \includegraphics[width=\linewidth]{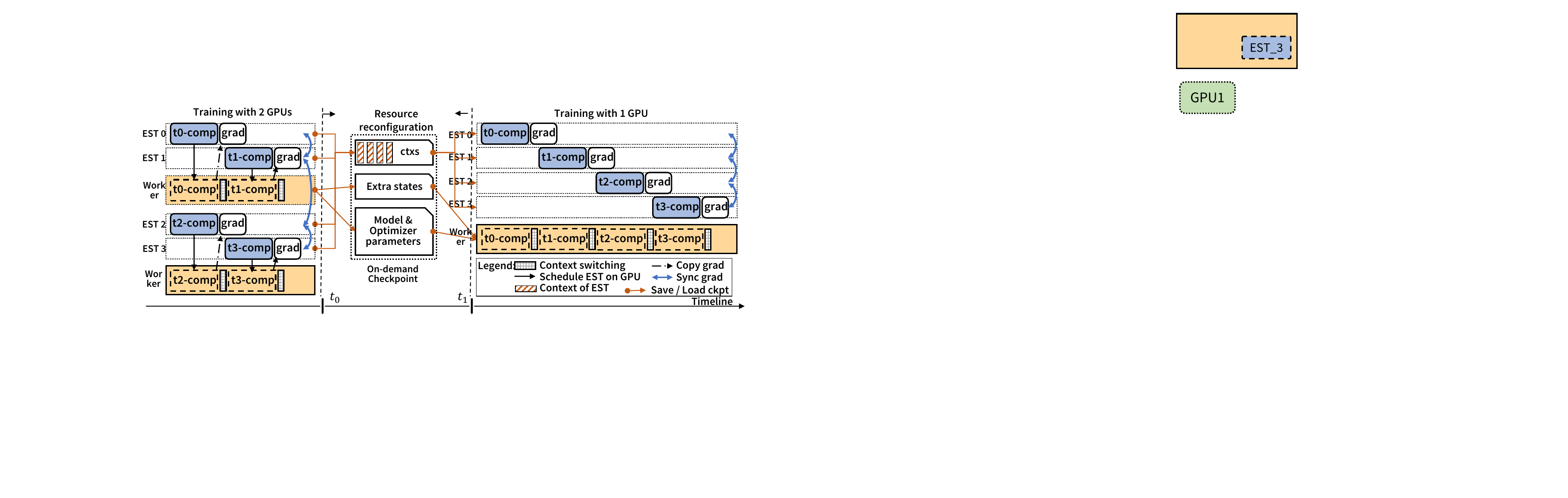}
        \caption{Execution flow of EasyScaleThread (EST).}
        \label{fig:easyscale-exec}
    \end{minipage}
\end{figure*}

The design principle of EasyScale has been inspired by the big data analysis systems~\cite{dean2004mapreduce,isard2007dryad,spark12,carbone2015apache,lin2016streamscope}, which guarantee the consistent output regardless of the allocated resources. Similarly, the DL frameworks that train neural network models by analyzing data samples, can be viewed as specialized data analysis systems for artificial intelligence, should also produce consistent model accuracy regardless of allocated resources. Previous elastic training frameworks fail to maintain the consistent model accuracy due to the changed behaviors of training procedure upon elastic GPU allocation. 

Different from existing approaches, we think elastic training should generate \textit{bitwise identical} model parameters compared to the non-elastic DDP training over a fixed number of GPUs. 
As a result, the model accuracy is also identical and consistent.
Figure~\ref{fig:overview} shows an example of elastic training by scaling in from four GPUs (\ref{fig:overview}(a)) to two GPUs (\ref{fig:overview}(b)). To preserve the training behaviors, ideally we would like the four training workers to be executed in parallel on two GPUs. However, by multiplexing multiple training workers on a GPU, the concurrent memory usage increases in the forward pass, which can easily lead to either out-of-memory (OOM) exceptions~\cite{zico2021} or significant overhead in memory swapping~\cite{antman2020, pipeswitch2020}.
Besides, the aggregated memory usage of CUDA contexts (including that of training framework and CUDA itself) is also considerable. 
For example, 16 workers on a 16GB V100 GPU costs 12GB GPU memory for CUDA contexts (around 750MB per context).

As illustrated in Figure~\ref{fig:overview}(c), the key challenges in achieving accuracy-consistent elastic training is to preserve the training behaviors (\eg, number of workers) as well as sharing GPU resources efficiently.
To address the above challenges in the design of \oursys{} framework, we introduce the abstraction of \ourth{} (EST) to decouple resource allocation from the training procedure (\S\ref{sec:easyscalethread}). 
We further split the states of a EST into a stateful context and a stateless part, and minimize the size of the context that needs to be saved and optimize the context switching overhead.
Then, we identify the sources of non-determinism across the training software stack and present our approach to eliminate non-determinism (\S\ref{sec:nondeterminism}). We also propose \oursys{} scheduler to better utilize homogeneous / heterogeneous GPUs regarding ESTs to improve cluster utilization (\S\ref{sec:policy}).

\subsection{EasyScaleThread}
\label{sec:easyscalethread}

We introduce \ourth{} (EST) as a key abstraction in \oursys{}.
EST is inspired by the classical \textit{thread} concept in operating systems
and the ``single program multiple data'' (SPMD) model adopted commonly in DL~\cite{barham2022pathways},
which can separate the training procedure from underlying resource allocation, and is flexible to enable resource sharing through context switching.
As shown in Figure~\ref{fig:overview}(c), each \oursys{} worker is launched on one GPU.
The execution of original DDP training workers (\eg, PyTorch workers) is treated as that of ESTs, and any EST (\ie, a thread) can be dynamically allocated to a \oursys{} worker (\ie, a process) during training.
In a \oursys{} worker, multiple ESTs take turns to occupy the GPU for training (\eg, model forward and backward passes) in the time-slicing manner.
\oursys{} hooks the key steps of model training, such as data loading, model backward, and model updating through users' annotations, therefore ESTs can perform efficient context switching at mini-batch boundaries.
The user-defined model training semantics, including model structure, data augmentation, batch size, \etc, remain as usual. 
Under the EST abstraction, users only need to consider the number of logical training workers for tuning the hyper-parameters (\eg, batch size and learning rate), which is the same as their experiences of using DDP on a fixed number of GPUs. However, with EST they can benefit from elastic training automatically without concerns about inconsistent accuracy. 

The execution of EST is similar to that of the worker of DDP training. For each training step, training samples are processed by conducting forward-backward computation over the current model to generate gradients as output. After gradient synchronization, the model is updated.
Figure~\ref{fig:easyscale-exec} illustrates the case when training with four ESTs, the available resource scales from two GPUs to one GPU. 
To enable efficient sharing among ESTs, each \oursys{} worker maintains one one CUDA context to share among ESTs, thus it does not consume multiple times of GPU memory.
For each global step of data-parallel training, the input data is split across all ESTs. During runtime, each \oursys{} worker schedules one EST at a time, executes it in the EasyScale worker by occupying one GPU (\ie{}, a local step), and all belonging ESTs gets executed in the time-slicing manner. The global step is completed after all ESTs finish execution of the local steps and all produced gradients are aggregated to update the model parameters.

\textbf{Context switching --} When context switching between ESTs, the training states of the ESTs need to be saved and swapped out from GPU to CPU in order to avoid over-subscribing GPU memory, which can be costly to deteriorate training throughput. 
The key to enabling lightweight context switching is to reduce the states to be saved.
\oursys{} leverages the unique characteristics of DL jobs to allow most states to be shared and reused among different ESTs. In addition, \oursys{} chooses to perform context switching of ESTs after the forward-backward computation at the boundaries of each mini-batch, which further reduces GPU-CPU data traffic.

Specifically, we propose the following two approaches to reduce the states to be saved during context switching: 
\textit{1)} locating the sources of non-determinism that affect the model accuracy and record only the necessary states (\eg, the states of random number generators (RNGs)),  
and \textit{2)} leveraging the data-parallel behavior of DDP training to minimize the working set for data swapping.
The working set resided in GPU memory of an EST can be classified into three categories, including temporal tensors and activations, model parameters and optimization states, and gradients~\cite{virtualflow2021}. During context switching, we handle each category differently to reduce the working set to be swapped to CPU side effectively.
Firstly, temporal tensors and activations are created in the forward pass and destroyed in backward pass after the gradient generation~\cite{gandiva2018, antman2020}. 
Their working set is automatically freed up at the end of mini-batches, which do not need to be swapped to CPU. Therefore, we constraint the minimal time slice of an EST's local step to one mini-batch.
Secondly, for the model parameters and optimizer states, 
a replica is maintained by each \oursys{} worker during training, and only updated at the end of global steps.
Therefore, they remain the same for all ESTs till all ESTs are finished, thus they can be reused among ESTs, with no need to be swapped to CPU.
Finally, the gradients are calculated based on the different input data across ESTs, and cannot be shared nor reused. Therefore, only the gradients need to be swapped to CPU during EST context switching.
Fortunately, the gradients are only used in distributed gradient synchronization at the end of a global step. 
To mitigate the cost of saving gradient working set, we overlap the gradient swapping with the backward computation of current EST and the forward computation of next EST to be switched in.
In such a way, each \oursys{} worker executes the ESTs alternately until all ESTs finish local steps. 
After that, the distributed gradient synchronization is triggered with model parameters updated.

\textbf{Adapting to elasticity --} When available resource (\eg{}, number of GPUs) of a training job changes (\aka, resource elasticity), \oursys{} adopts on-demand checkpointing to preserve necessary states, as shown in Figure~\ref{fig:easyscale-exec}.
The checkpoint contains the contexts of all ESTs, the extra states (including the training progress and other states for achieving accuracy-consistency illustrated in \S\ref{sec:nondeterminism}), and parameters (\eg{}, model, optimizer, and learning rate scheduler).
Unlike the EST contexts, only one replica of the extra states and parameters is required, because they can be shared across ESTs within each global training step.
Note that after continuing the model training on the reconfigured resources, 
\oursys{} manages each worker loads a copy of extra states and parameters, as well as the corresponding contexts of re-distributed ESTs, so that all ESTs can resume from the last saved states.

\textbf{Optimizing data pre-processing --} Since the data pre-processing is becoming more resource-intensive, existing DL frameworks such as PyTorch commonly use standalone data workers on CPUs to accelerate the model training on GPUs. Specifically, the data workers run asynchronously to the training workers for data supply, load samples and perform data augmentations to build training batches.
To further optimize the training efficiency with ESTs, we need to consider the data supply along with EST execution.
The number of data workers is usually configured for each training worker by users to ensure the GPU training is not being blocked by data supply. 
Naively scaling the number of data workers regarding the number of ESTs can easily lead to massive CPU processes, thus overwhelming the training system. For example, if the users configure 8 data workers per training worker, for a case with 16 ESTs sharing a GPU, the total data workers on one machine can be 128. 
In \oursys{}, we can avoid the above problem by sharing data workers among all ESTs, because only one EST is executing within a \oursys{} worker at any time. Therefore, the data consuming rate is similar to that of executing one training worker. 

\begin{figure}[t]
    \centering
    \includegraphics[width=0.8\linewidth]{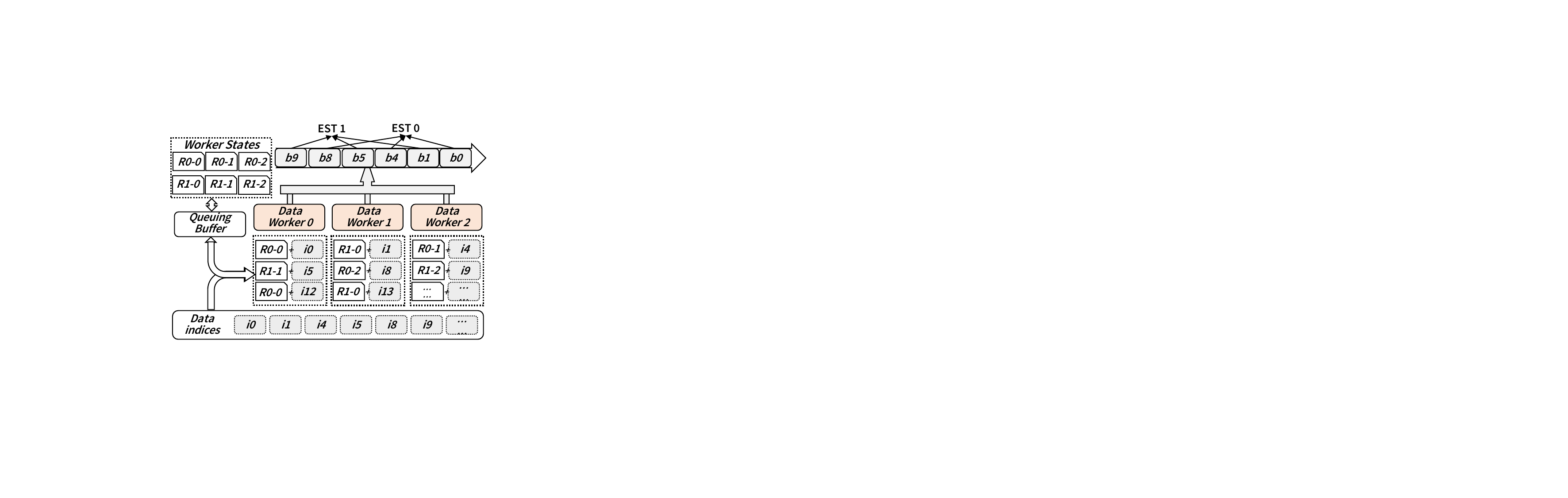}
    \caption{Data worker sharing.}
    \label{fig:dataloading}
\end{figure}

To enable data worker sharing, \oursys{} employs a distributed data sampler that jointly considers the global indices of ESTs and the time-slicing manner to generate data indices in a queue.
The data indices are then processed by data workers in order.
Figure~\ref{fig:dataloading} shows the case of sharing three data workers between two ESTs, 
where the total number of ESTs is four (\ie{}, training with two GPUs shown in Figure~\ref{fig:easyscale-exec}).
The training batches of $EST 0$ and $EST 1$ are \textit{b0} and \textit{b1} for mini-batch 0, and \textit{b4} and \textit{b5} for mini-batch 1.
The state of data worker \textit{j} to process data indices for EST \textit{i} on a dedicated GPU is denoted as \textit{Ri-j}, shown in Figure~\ref{fig:dataloading}.
Note that due to the asynchronous execution of data workers, the progress of data workers (\eg{}, mini-batch index) is usually ahead of the training progress. 
To maintain the consistent state for elastic training, a queuing buffer is introduced to record the necessary states (\eg{}, the state of RNG) for mini-batches that ESTs do not consume. 
To balance the load, data workers in \oursys{} take turns to obtain the corresponding state (\eg{}, \textit{Ri-j}) of given data indices from the queuing buffer for pre-processing, 
and the state is committed back to the queuing buffer once pre-processing is finished. 
Since the worker states are dequeued from the queuing buffer according to training progress, they are categorized as an extra state during the on-demand checkpointing. 

\subsection{Eliminating Non-determinism}
\label{sec:nondeterminism}

While ESTs can decouple the resource allocation from the training procedure and facilitate efficient resource sharing through EST context switching, using ESTs still may result in inconsistent accuracy compared to the DDP training.
Besides, utilizing elastic training to make use of heterogeneous GPUs can also introduce non-trivial, non-deterministic, and previously unstudied behaviors. 
To tackle non-determinism during training, we use a top-down approach comparing the tensors of \oursys{} and DDP. We conduct experiments using the same number of workers but with different configurations to identify the factors that impact training accuracy in bitwise.
Surprisingly, we find that the root causes are scattered across almost the entire software stack used for training. 

For the \textit{implicit framework state}, existing training frameworks commonly maintain several implicit states, which must be consistent throughout the training for determinism.
Although they organize operators (\eg, convolution, batch normalization) in a computation graph~\cite{abadi2016tensorflow}, several operators implicitly rely on additional states beyond their predecessors' outputs.
For instance, the \texttt{Dropout} operator depends on the random number generator (RNG) states, whereas the \texttt{BatchNorm} operator tracks its running states. In addition, the data loader and data augmentation also depend on RNG states from Python, NumPy, PyTorch, \etc{}

For the \textit{communication mechanism}, the gradient synchronization via all-reduce used in DDP is non-deterministic when resource elasticity is involved.
During synchronization, the gradients are gathered into gradient buckets to achieve higher throughput and lower latency~\cite{pytorch-distributed}. 
The mapping of gradients to buckets firstly follows a static reversed topological order of the computation graph, and then is reconstructed at the end of the first mini-batch based on the order of when gradient tensors are derived. 
However, when resource scales in/out, the training workers will restart and rebuild the communication channels, which changes the mapping and eventually disrupts the gradient aggregation order, leading to non-determinism together with the all-reduce implementation~\cite{nccldeterministic, frameworkdeterminism}. 

For the \textit{Operator implementation}, existing training frameworks may select different implementations for the same operator during training, which can result in subtle differences in training accuracy.
There are two reasons why different operator implementations are selected. 
Firstly, profiling-based optimizations adopted by frameworks~\cite{torchcudnnbenchmark}, compilers~\cite{sivathanu2019astra}, or vendor libraries~\cite{cudnnselect} can apply various kernel implementations on GPUs to optimize operator performance based on profiling results across mini-batches. Secondly, the kernel implementation may be hardware-specific, such as implementations designed for a specific number of SM units and low-bit components, which makes it unsuitable for all types of GPUs.

\textbf{Solutions for different levels of determinism --} To address the non-determinism across the software stack, \oursys{} defines different levels of determinism for elastic training, and applies solutions to guarantee consistent accuracy for each level.

\textit{D0: Static determinism} -- Multiple training runs with a fixed number of GPUs should always result in identical model accuracy.
To achieve \textit{D0}, consistent framework states and operator implementations are required. 
As for consistent framework states, we fix the random seeds of RNGs at the beginning of training and record the RNG states of both the data workers and ESTs in the extra states and EST contexts of the on-demand checkpoint. 
As for operator implementations, we disable profiling-based optimizations and select deterministic kernel implementations (\eg{}, without atomic instructions). DL frameworks such as PyTorch and TensorFlow also recommend this approach to improve model reproducibility~\cite{pytorchreproducibility, frameworkdeterminism}.

\textit{D1: Elastic determinism} -- Multiple training runs with different numbers of GPUs should always result in identical model accuracy.
\textit{D1} requires necessitates resolving the non-deterministic aspects of the communication mechanism beyond \textit{D0}.
To achieve this, we assign a constant virtual communication rank to each EST and store the indices that make up the gradient buckets in the checkpoint. 
When resource scales in/out, the training recommences using the checkpoint and reconstructs the gradient buckets by initially reinstating recorded indices of gradient-bucket mapping. Subsequently, reconstruction of the communication channel is disabled.

\textit{D2: Heterogeneous determinism} -- Multiple training runs with different types of GPUs should always result in identical model accuracy.
To achieve \textit{D2}, we develop hardware-agnostic operator implementations on GPU, involving two main aspects: \textit{1)} we modify operator implementations (\eg, \texttt{reduce}, \texttt{dropout} in PyTorch) by selecting a specific number of SMs and threads that can run on any type of GPU, 
and \textit{2)} we deterministically choose the same operator implementations (\eg, \texttt{convolution} in cuDNN, and \texttt{gemm}, \texttt{gemv} in cuBLAS) by fixing the algorithm identifier (\textit{algo\_id}) in library calls.

In \oursys{}, \textit{D0} and \textit{D1} are enabled by default due to their negligible overhead (\S\ref{sec:eval_determinism}). 
However, enabling \textit{D2} may result in noticeable overhead for certain types of operators such as convolution, because they cannot utilize vendor-optimized kernels on GPU.
To address this issue, \oursys{} automatically analyzes a DL model by scanning the PyTorch \texttt{nn.Module} to identify whether it relies on hardware-specific kernel optimizations. If not, \textit{D2} is enabled and elastic training can use heterogeneous GPUs. Otherwise, \oursys{} restricts its use to homogeneous GPUs.

EasyScale does not rewrite operators, it only selects deterministic operator implementations or specifies allowed SMs/threads for existing operators. Currently, EasyScale has already supported diverse models (CV/NLP models in Table~\ref{tab:trace_workloads}), popular structures (e.g., \texttt{convolution}, \texttt{transformer}), and fundamental operators (e.g., \texttt{gemm}, \texttt{reduction}). In the future, we will extend EasyScale to support more deterministic operators, therefore less effort is needed from the users to achieve \textit{D2} determinism. Notably, to achieve operator determinism, EasyScale is analogical to prior works (e.g., Rammer~\cite{rammer}, REEF~\cite{REEF}) that accelerate training process through novel operator designs. Currently, hardware-specific kernels can be supported in EasyScale with \textit{D1} determinism. For future work, we will allow the users to customize \textit{D2} kernels via Cutlass to enable the exploration of more hardware features for heterogeneous GPUs.

\subsection{\oursys{} Scheduler}
\label{sec:policy}

\begin{figure}[t]
    \centering
    \includegraphics[width=0.99\linewidth]{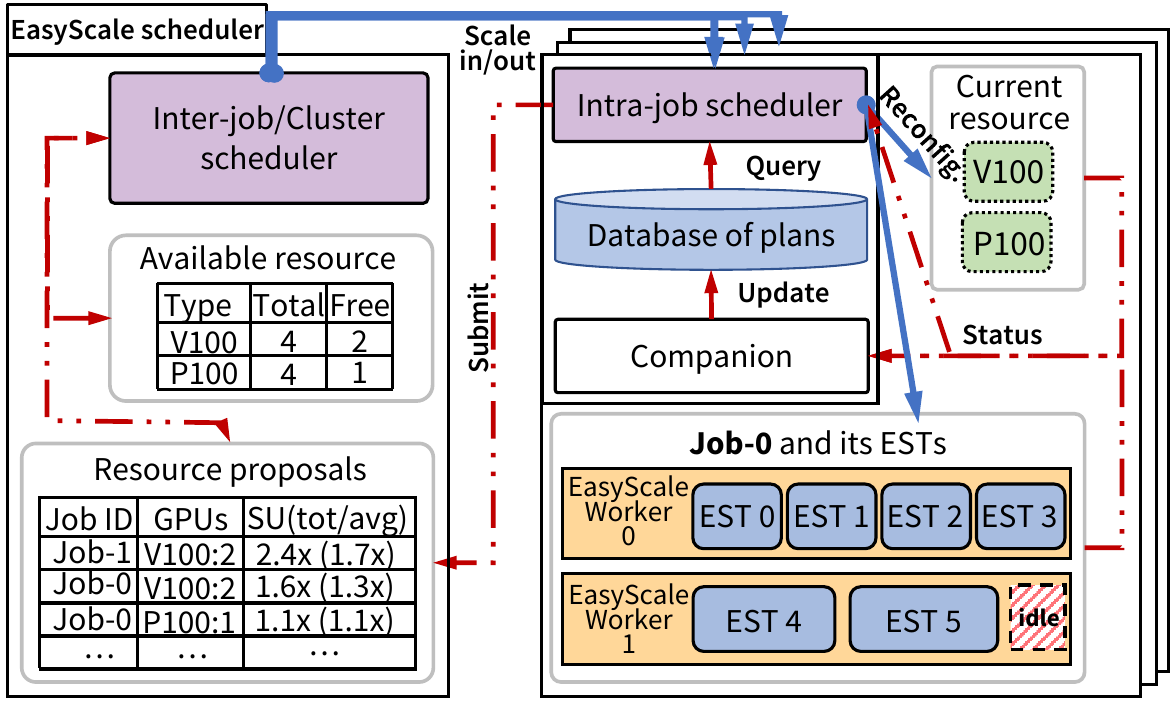}
    \caption{Workflow of the \oursys{} scheduler.}
    \label{fig:easyscale-scheduler}
\end{figure}

In this section, we describe the \oursys{} scheduler, which is co-designed with the \oursys{} framework, to improve the training job throughput and the cluster utilization, guaranteeing of consistent accuracy at the same time. 
It is capable of scheduling the \oursys{} jobs to available GPUs of the cluster. 
For each job, it can schedule each job to a set of heterogeneous GPUs, utilizing the \oursys{} framework's capability of decoupling training procedure from underlying resources.
The \oursys{} scheduler has a hierarchical architecture, as illustrated in Figure~\ref{fig:easyscale-scheduler}. The \textbf{inter-job (cluster) scheduler} is responsible for allocating resources among jobs at the cluster scale. Additionally, each training job has an \textbf{intra-job scheduler} that coordinates ESTs and currently allocated GPUs.

In \oursys{}, the scheduling decision is dispatched from the inter-job (cluster) scheduler to the intra-job scheduler. Then, the intra-job schedulers of current jobs reconfigure (\ie{}, scale in/out) the occupied GPUs resources respecting the scheduling decision, query the database to get the best plan, and consequently reschedule the ESTs to current resources. 
In contrast, the scheduling decision comes from the resource proposals submitted by the intra-job schedulers, and is approved by the inter-job scheduler. Specifically, each intra-job scheduler leverages \textbf{a standalone companion module} to maintain the database of plans.
And the intra-job scheduler tries to scale out with incremental homogeneous GPUs (\eg{}, 2 V100), thereby selecting top-K plans according to estimated throughput improvement as the resource proposals to submit to the cluster scheduler.
The companion module leverages the job information reported by \oursys{} framework to initialize the database and actively update the database once it has monitored significant biases between the estimated throughput and the reported throughput. 

\textbf{Intra-job scheduler --}
Its main responsibility is to generate the EST-to-GPU mapping configurations with the help of the companion module and the database of scheduling plans. There are three roles for the intra-job scheduler.
\textit{Role-1:} Under current available GPU resources, it queries the database and applies the top-1 configuration with the highest estimated throughput.
\textit{Role-2:} Supposing the job is allowed to scale out for higher training throughput, it queries the database to explore new configurations, calculates the incremental GPU number and the estimated speedup, and then forms the resource proposals submitted to the inter-job scheduler.
\textit{Role-3:} Once it receives any scheduling plan, it scales in/out the GPU resources accordingly and intermediately. Then it reschedules the ESTs to current GPUs (\textit{Role-1}) and generates the resource proposals later (\textit{Role-2}). Additionally, if it has observed the slowdown with incremental resources, it falls back to using previous resources and releases the newly allocated ones.

\textbf{Companion module --}
The intra-job scheduler's companion module keeps a database of scheduling plans that consider the available GPU types and the maximum number of ESTs ($maxP$) specified by DL developers during model designing stage. Each plan includes the quantity of GPU resources, EST-to-GPU mapping configuration, and estimated throughput based on job details and performance models. When a job runs for the first time, the companion module initializes the database using historical data~\cite{weng2022mlaas}.

The companion module aims to achieve load balance among different GPUs by generating plans that allocate proper ESTs, based on the quantum property of EST allocation (\ie{}, integer number) and the consecutiveness of GPUs' computing capability. To estimate the throughput of EST-to-GPU mapping configurations, it uses an analytical performance model. Specifically, it introduces a new metric called \textit{waste} to indicate the quantity of wasted computing capability due to load imbalance, which can represent two scenarios: \textit{1)} allocated ESTs cannot fully utilize GPUs' computing capability (Equation~\ref{equ:waste_1}--\ref{equ:waste_2}), and \textit{2)} the EST allocation is over-provisioned to satisfy the $\#\mbox{\textit{of total ESTs}} \ge maxP$ constraint (Equation~\ref{equ:waste_0}). The number of available GPUs is denoted as $N_i$, where $i$ represents GPU type. The workload-related computing capability $C_i$ is estimated as mini-batches per second, and the maximum number of assigned ESTs for GPU type $i$ is denoted as $A_i$. Additionally, an overload factor ($f_{overload}$) represents the maximum overload for requested GPU types. If one GPU type undertakes too many ESTs, it becomes a bottleneck and slows down other GPUs due to Sync-SGD. Finally, we derive estimated throughput by subtracting \textit{waste} from our calculations (Equation~\ref{equ:waste_3}).

\begin{subequations}
    \fontsize{7.5}{7.5}\selectfont
    \label{equ:waste}
    \begin{align}
        &\ nEST = \sum_{i} N_i \times A_i, \mbox{ \ $nEST \ge maxP$} \label{equ:waste_0}\\
        &f_{overload} = \max_{i, N_i > 0} {A_i}/{C_i} \label{equ:waste_1}\\
        &waste = \sum_{i, N_i > 0} \big( C_i - {A_i}/{f_{overload}} \big) + \big(\ nEST - maxP\big) / f_{overload} \label{equ:waste_2}\\
        &throughput = (\sum_{i} N_i \times C_i) - waste \label{equ:waste_3} 
    \end{align}
\end{subequations}

\textbf{Inter-job cluster scheduler --}
It evaluates the submitted proposals by considering available resources and proposal priorities.
To improve the aggregated job throughput and cluster utilization, it adopts a heuristic (greedy policy in practice) that tends to accept the proposals with a higher speedup per GPU. If multiple proposals offer the same speedup, it prioritizes the one with more GPUs. 
By synchronizing fluctuating free resources to the table of available resources, it supports co-locating \oursys{} jobs with other non-elastic jobs (such as online serving jobs), making optimal use of temporarily available idle resources.
Notably, the inter-job scheduler reserves flexible interfaces for users to experiment with other scheduling policies. If needing more contexts for scheduling, the intra-job scheduler can also be easily extended to report more framework/resource information to inter-job scheduler.


\section{Implementation}
\label{sec:impl}

The \oursys{} framework is built on PyTorch 1.8 LTS ~\cite{paszke2019pytorch} and requires approximately 1,200 lines of Python code and 2,000 lines of C++ modifications to PyTorch. The C++ implementation includes a distributed data-parallel communication library called \texttt{ElasticDDP}, which supports communication among multiple ESTs for all-reducing gradients and building communication buckets consistently during resource elasticity. Execution control flow and context switching are implemented as an add-on PyTorch module.

A prototype cluster scheduler of the \oursys{} scheduler is implemented on Kubernetes~\cite{k8s} for evaluation. We implement AIMaster, which includes the intra-job scheduler and the companion module, with around 2,000 lines of Python code. AIMaster performs three main functions: collecting performance profiling reported by \oursys{} runtime through an RPC library; submitting resource proposals; monitoring resource allocation timeout through a Kubernetes Python informer; and containing a policy controller to calculate and submit incremental resource requests to the cluster scheduler.
We adopt on-demand checkpoint~\cite{gandiva2018} to record DL model, epoch, and necessary states mentioned in \S\ref{sec:nondeterminism} to support continuous job training when resource elasticity occurs.

The \oursys{} jobs run in Docker containers with \oursys{} installed within our internal GPU cluster scheduler that is optimized from Kubernetes version. Currently, we have deployed \oursys{} on two internal GPU production clusters used for serving DL development (\ie{}, Jupyter Notebook) and model inference respectively. One deployed cluster consists of more than 10K GPUs.


\section{Evaluation}
\label{sec:eval}


In this section, we present the evaluation of \oursys{}. Firstly, we demonstrate its accuracy-consistency and efficiency through micro-benchmarks. Secondly, we evaluate it using real workloads on a small cluster with 64 GPUs to show its scheduling efficiency. Lastly, we evaluate it on a production cluster equipped with thousands of GPUs to highlight the improvement in cluster utilization.

The micro-benchmark experiments and trace experiments are conducted on a cloud GPU cluster with 16 servers, specially, 4 servers each with 8 V100, 8 servers each with 2 P100, and 4 servers each with 4 T4. 
Each server runs CentOS 7.8, and their GPUs are powered by Nvidia driver 450.102.04, CUDA 10.1, and CuDNN 7.
As for the workloads, eight state-of-the-art DL models are selected from Github, together with open datasets, as summarized in Table~\ref{tab:trace_workloads}. 
They are implemented based on PyTorch 1.8 LTS, and are ported to \oursys{} with a few lines of code changing.

\begin{table}[t]
    \centering
    \small
    \caption{Deep learning workloads in experiments.}
    \begin{tabular}{|l|l|l|}
        \hline 
        Model            & Task                   & Dataset    \\ \hline
        ShuffleNetv2~\cite{ma2018shufflenet}     & Image   Classification & ImageNet~\cite{imagenet09}   \\ \hline
        ResNet50~\cite{resnet16}         & Image   Classification & ImageNet~\cite{imagenet09}   \\ \hline
        VGG19~\cite{vgg14}            & Image   Classification & ImageNet~\cite{imagenet09}   \\ \hline
        YOLOv3~\cite{yolov318}           & Object Detection       & PASCAL~\cite{everingham2010pascal} \\ \hline
        NeuMF~\cite{he2017neural}            & Recommendation         & MovieLens~\cite{harper2015movielens}  \\ \hline
        Bert~\cite{bert19}             & Question Answering   & SQuAD~\cite{squad16}      \\ \hline
        Electra~\cite{electra20}          & Question Answering     & SQuAD~\cite{squad16}      \\ \hline
        SwinTransformer~\cite{swintransformer2021} & Image   Classification & ImageNet~\cite{imagenet09}   \\ \hline
    \end{tabular}
    \label{tab:trace_workloads}
\end{table}

\subsection{Micro-benchmarks}
\label{sec:benchmark_exp}

\subsubsection{Ensuring accuracy-consistency} \quad \par

To demonstrate the accuracy-consistency of \oursys{}, which guarantees to produce bitwise-identical DL models under an elastic number of heterogeneous GPUs, 
we use \oursys{} to train the DL workloads of Table~\ref{tab:trace_workloads} in three different stages with different GPU configurations: 
\textit{stage 0} with 4 V100 GPUs, \textit{stage 1} with 2 V100 GPUs, and \textit{stage 2} with 1 V100 and 2 P100 GPUs. 
Changing from from \textit{stage 0} to \textit{stage 1} represents the \textit{resource elasticity}, and changing from from \textit{stage 1} to \textit{stage 2} represents the \textit{resource heterogeneity}.
In each stage, the workloads are trained for 100 mini-batches.

\begin{figure}[htbp]

    \subfigure{
        \centering
        \includegraphics[width=0.46\linewidth]{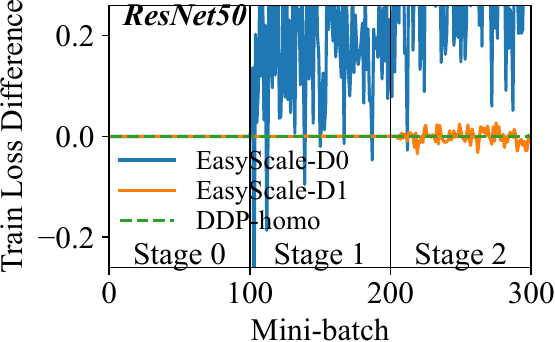}
    }
    \vspace{-0.1in}
    \hfill
    \subfigure{
        \centering
        \includegraphics[width=0.46\linewidth]{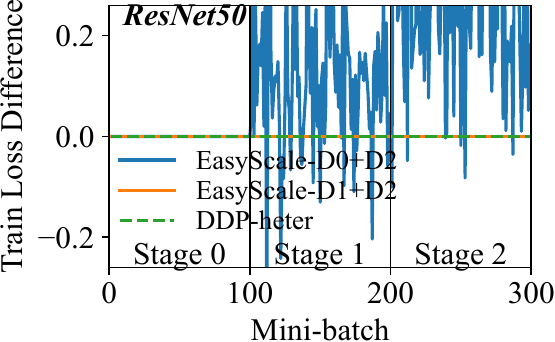}
    }
    \vspace{-0.1in}
    \setcounter{subfigure}{0}
    \subfigure[Under homogeneous determinism configurations]{
        \centering
        \includegraphics[width=0.46\linewidth]{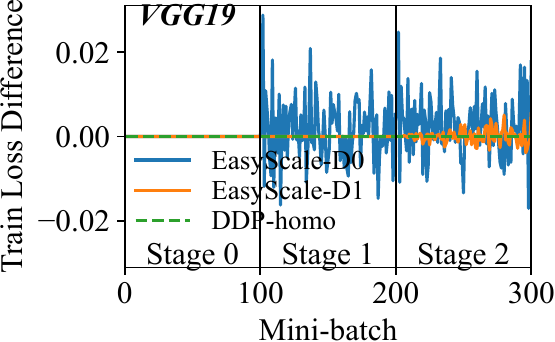}
    }
    \hfill
    \subfigure[Under heterogeneous determinism configurations]{
        \centering
        \includegraphics[width=0.46\linewidth]{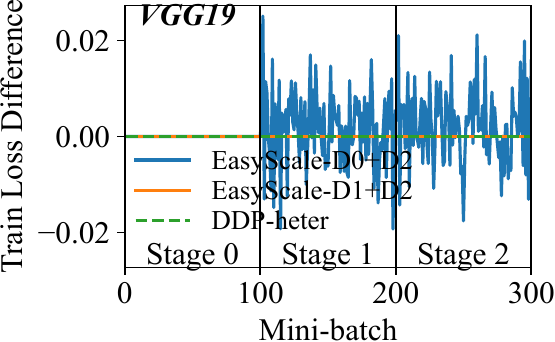}
    }
    \caption{Loss curve difference of EasyScale and DDP. }
    \label{fig:accuracy}
\end{figure}

We use PyTorch DDP with 4 V100 GPUs as the baselines. 
Both \oursys{} and DDP have 4 workers in total (\ie{}, 4 ESTs for \oursys{}).
\oursys{} is configured with four determinism configurations: two homogeneous determinism configurations (\textit{D0} and \textit{D1}) and two heterogeneous ones (\textit{D0+D2} and \textit{D1+D2}).
DDP has two corresponding configurations, \textit{DDP-homo} with fixed random seeds and deterministic algorithms to ensure the reproducibility, and \textit{DDP-heter} with additional selection of heterogeneous deterministic kernels (originally belong to \textit{D2}).
Figure~\ref{fig:accuracy} shows the loss curve differences of the last worker on \texttt{ResNet50} and \texttt{VGG19}. 
The train loss of \textit{D1} is identical to that of \textit{DDP-homo} in \textit{stage 0} and \textit{stage 1}, and the train loss of \textit{D1+D2} is identical to that of \textit{DDP-heter} in all stages, demonstrating how \oursys{} can preserve consistent accuracy.

By comparing the curves of \textit{D0} with \textit{D1}, and also \textit{D0+D2} with \textit{D1+D2}, we can highlight the elasticity determinism of \oursys{}. We have observed that both \textit{D0} and \textit{D0+D2} start experiencing loss differences since \textit{stage 1} after checkpointing and restarting. This is because \textit{D0} ignores the states of gradient-to-bucket mapping in the checkpoint, which results in losing these states after restarting. In contrast, \textit{D1}/\textit{D1+D2} records these states in the checkpoint and thus have identical loss curve to \textit{DDP-homo}/\textit{DDP-heter}.

Furthermore, by comparing \textit{D1} with \textit{D1+D2}, we can highlight the heterogeneous determinism.
Specifically, in \textit{D1}, loss differences begin to emerge from \textit{stage 2}, due to automatic selection of different low-level kernel implementations on heterogeneous GPUs.
However, enabling \textit{D2} to fix the kernel selection for EasyScale and DDP eliminates loss difference in \textit{stage 2}.
The results of the other models are similar and have been omitted due to space constraints.
In summary, EasyScale with \textit{D1+D2}, can ensure the accuracy-consistency with DDP after any number of training iterations.

\subsubsection{Efficient GPU resource sharing} \quad \par

To demonstrate EasyScale's ability to efficiently run multiple workers (ESTs) on a single GPU, we compared it with worker packing proposed by Gandiva~\cite{gandiva2018}. Worker packing involves multiplexing the same GPU across multiple workers and is another potential method for achieving accuracy-consistency with resource elasticity. We trained two typical models, \texttt{ResNet50} and \texttt{ShuffleNetV2}, on a V100 GPU. The batch size of \texttt{ResNet50} is set to 32 as it is commonly used in benchmarks. The batch size of \texttt{ShuffleNetV2} is set to 512 to fully utilize the 32GB V100 memory using one worker, which is typically how DL researchers utilize a GPU's capability.
EasyScale is configured as \textit{EasyScale-D1}, and worker packing is implemented with \textit{DDP-homo} to ensure the reproducibility.

\begin{figure}[htbp]
    \subfigure[ResNet50]{
        \includegraphics[width=0.46\linewidth]{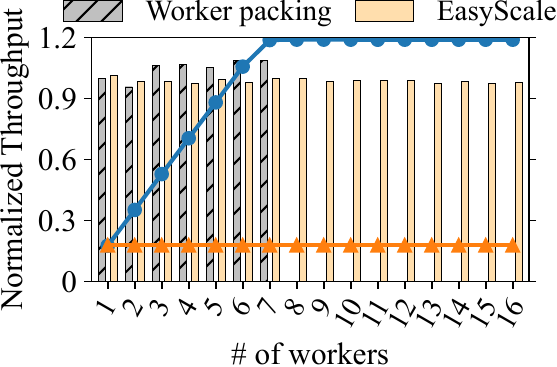}
        \vspace{-0.1in}
        \label{fig:exp_advantage_resnet}
    }
    \subfigure[ShuffleNetv2]{
        \includegraphics[width=0.46\linewidth]{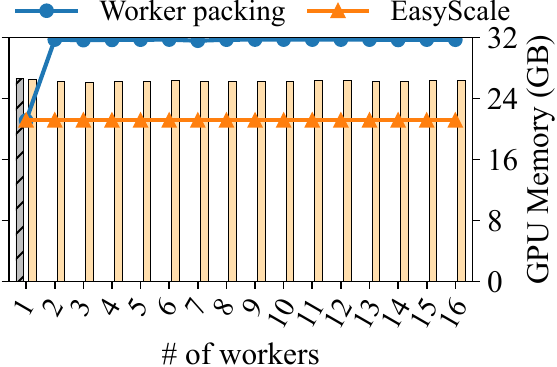}
        \vspace{-0.1in}
        \label{fig:exp_advantage_shufflenet}
    }
    \vspace{-0.1in}
    \caption{The peak GPU memory usage (curves) and throughput (bars) of \oursys{} and worker packing when executing multiple ESTs/workers on a V100.}
    \label{fig:exp_advantage}
\end{figure}

We conducted 10 runs of \oursys{} and worker packing with varying numbers of workers. Figure~\ref{fig:exp_advantage} shows the training throughput (batch size divided by average mini-batch time) and peak GPU memory usage. All throughput values are normalized to one worker under worker packing.
As expected, when running only one worker, both methods have similar throughput and memory usage. However, as the number of workers increases, the GPU memory usage for \oursys{} remains constant while worker packing experiences a gradual increase in GPU memory usage. Worker packing suffers from out-of-memory (OOM) exceptions after 8 workers for \texttt{ResNet50} and 2 workers for \texttt{ShuffleNetV2}.
In contrast, \oursys{} carefully reuses the DL components across ESTs such as model parameters and optimizer states while minimizing EST context. As a result, its GPU memory usage remains almost constant regardless of the number of workers.
Furthermore, \oursys{} has an almost constant training throughput regardless of the worker number. But worker packing grows at the beginning and reaches $1.11\times$ compared to \oursys{}, resulting from higher GPU utilization due to the concurrent execution of multiple kernels, but at a cost of higher memory usage as shown above. 

\begin{figure}[htbp]
    \centering
    \includegraphics[width=\linewidth]{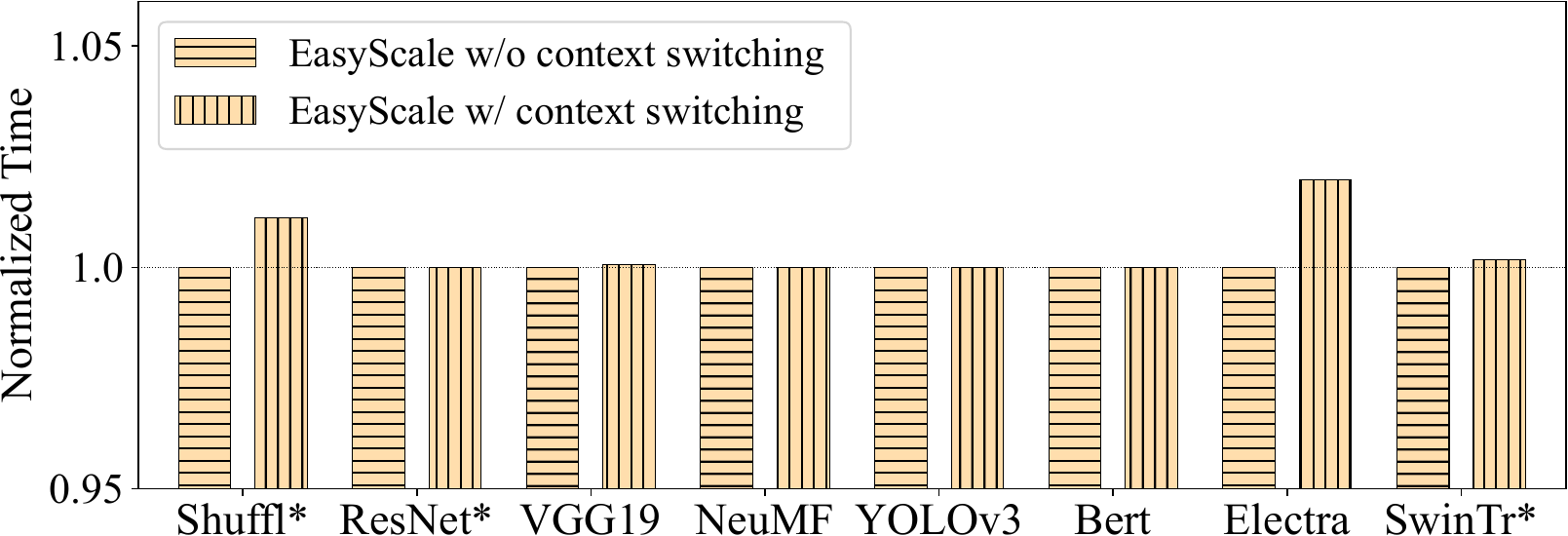}
    \caption{The time of lightweight context switching.}
    \label{fig:exp_context_switch}
\end{figure}

To demonstrate the lightweight context switching, we run different workloads using one EST per GPU, with and without the context switching. Note that \oursys{} cannot generate accuracy-consistent results same as DDP when there is no context switching. 
Figure~\ref{fig:exp_context_switch} illustrates that in most cases, the overhead is negligible, with a maximum of 1.9\% for \texttt{Electra}, because \oursys{} meticulously identifies non-determinism and only records determinism-critical states instead of large model parameters in contexts.

We further evaluate the data worker sharing optimization among above workloads with 8 ESTs. 
Enabling this optimization results in an average decrease of 67.1\% in training time for the first mini-batch. This is because data worker sharing significantly reduces the number of required data workers (\eg{}, reduced from 32 to 4), thereby reducing their launch time when responding to elasticity.

\subsubsection{Overhead of ensuring accuracy-consistency} \quad \par
\label{sec:eval_determinism}

\begin{figure}[t]
    \centering
    \includegraphics[width=\linewidth]{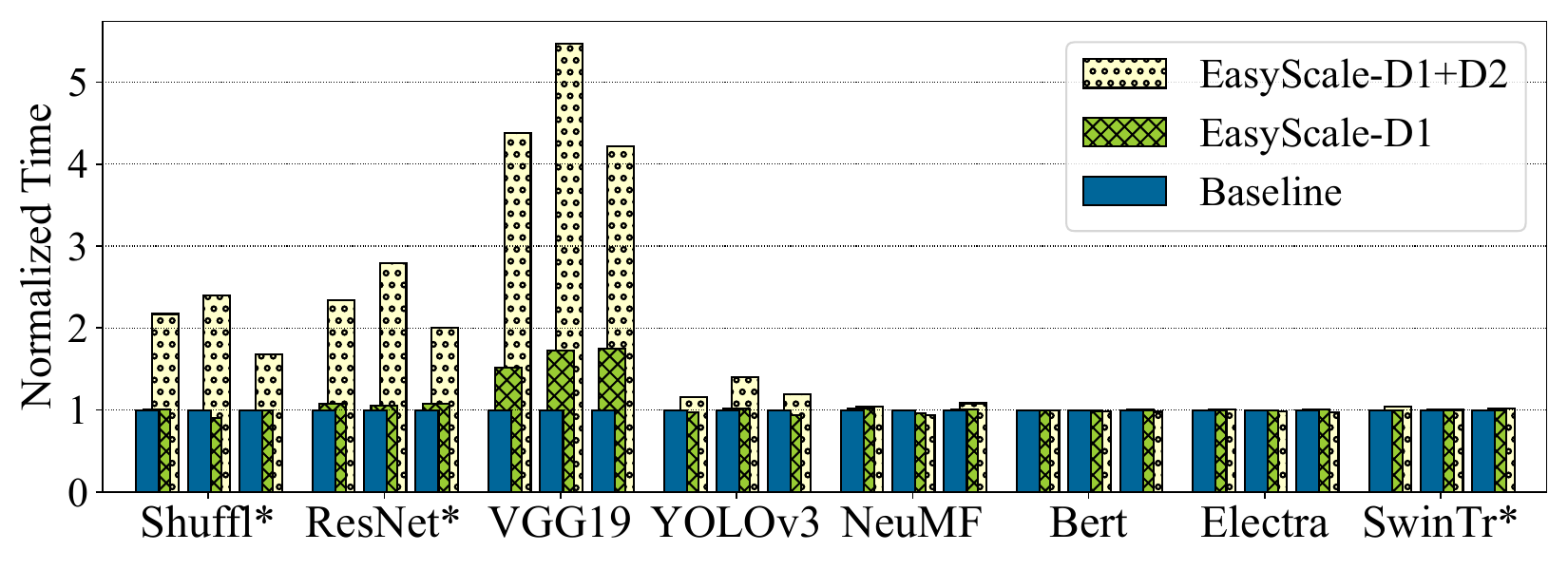}
    \caption{
        The overhead of ensuring accuracy-consistency.
        Each bar reports the time normalized to the baseline of the specific GPU. 
        For each model, the three bars indicate the results in V100, P100, and T4, from left to right. 
    }
    \label{fig:exp_overhead_heterogeneous}
\end{figure}

We examine the overhead of ensuring accuracy-consistency by measuring the training time of typical DL workloads.
EasyScale is configured with two configurations: \textit{a)} \textit{D1}, which ensures accuracy consistency when using an elastic number of homogeneous GPUs, and \textit{b)} \textit{D1+D2}, which ensures consistency when using an elastic number of heterogeneous GPUs. 
The baseline is set as the official version of PyTorch. The experiment is conducted on V100, P100, and T4 GPUs. 
Figure~\ref{fig:exp_overhead_heterogeneous} presents the per-iteration time normalized to the baseline for each type of GPU. 

The models can be classified into two categories based on their overhead.
The first category includes models such as \texttt{NeuMF}, \texttt{Bert}, \texttt{Electra}, and \texttt{SwinTransformer}. For these models, ensuring accuracy consistency (including both \textit{D1} and \textit{D1+D2}) results in less than 1\% overhead. Therefore, we can train them using elastic and heterogeneous GPU resources with negligible overhead.
The second category includes models such as \texttt{ShuffleNetV2}, \texttt{ResNet50}, \texttt{VGG19}, and \texttt{YOLOv3}. Ensuring consistency on homogeneous GPUs for these models also brings negligible overhead. However, ensuring consistency on heterogeneous GPUs will introduce considerable overhead (\ie{}, 236\% on average). This is because \oursys{}-\textit{D2} turns off vendor-optimized convolution kernels in these workloads for determinism.
Nevertheless, \oursys{} can automatically identify the training jobs that do not rely on such kernels and allow them to use elastic and heterogeneous GPU resources while using homogeneous GPU resources for other jobs instead.

\begin{figure}[htb]
    \centering
    \includegraphics[width=\linewidth]{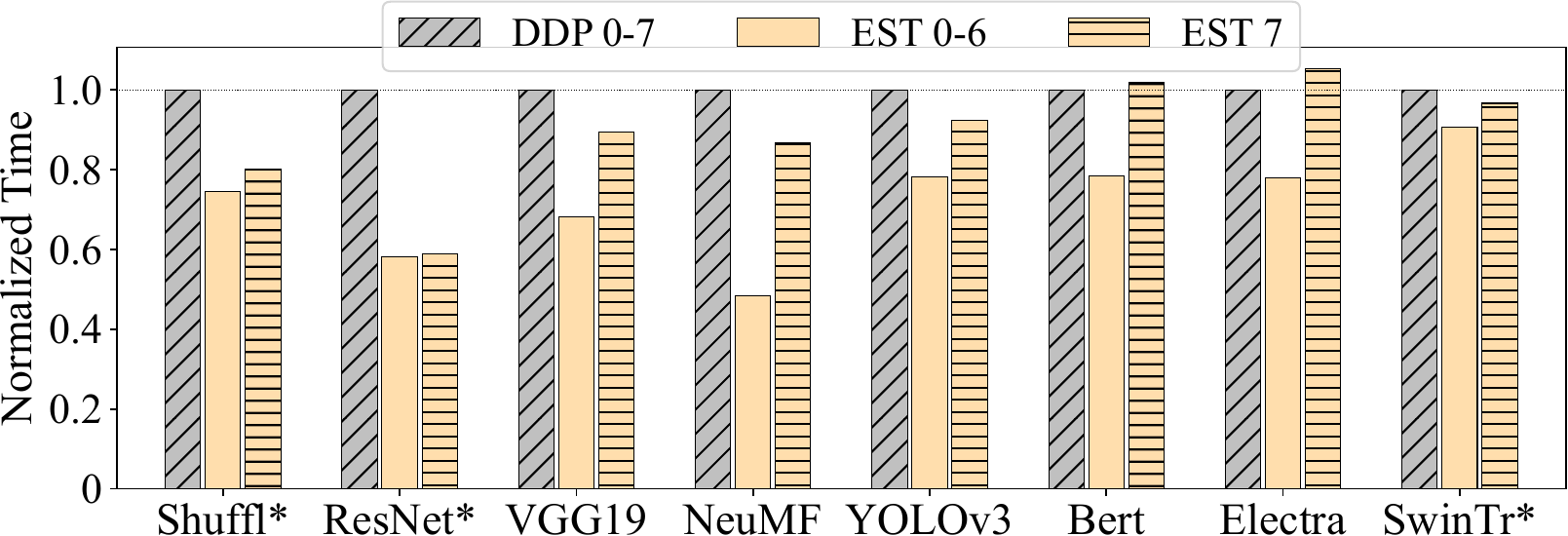}
    \caption{The overhead of gradients copy and sync.}
    \label{fig:exp_gradient_worker}
\end{figure}

We further measure the gradient copy and synchronization overhead that the EST abstraction might introduce.
\oursys{} is configured to execute 8 ESTs on 1 GPU, while DDP runs on 8 GPUs.
To ensure accurate results, we skip the first 10 mini-batches for warm-up and recorded the average execution time of each EST. EST 0-6 asynchronously copies the generated gradients through D2H operations, and EST 7 performs gradient synchronization similar to DDP.
Surprisingly, as shown in Figure~\ref{fig:exp_gradient_worker}, \oursys{} achieves superior or competitive performance compared to DDP. 
For EST 0-6, this is because of the overlapping between gradient copy and the backward computation as well as the forward computation of next EST.
For EST 7, this is because when EST 7 starts gradient synchronization, the other replicas of gradients (EST 0-6) are already ready for synchronization. 
In contrast with our findings in \oursys{}, it is difficult to ensure simultaneous production of gradients among all workers in DDP, which could lead to potential delays. 
Besides, when only one EST resides on each GPU, the gradient copy is not needed, and \oursys{} shows competitive performance to DDP.

\subsection{Trace Experiment}
\label{sec:trace_exp}

To demonstrate the improved resource utilization and job throughput of \oursys{}, we conducted trace experiments on a cloud cluster consisting of 32 V100 GPUs, 16 P100 GPUs, and 16 T4 GPUs.

\begin{figure*}[t]
    \begin{minipage}[t]{0.32\linewidth}
        \centering 
        \includegraphics[width=0.9\linewidth]{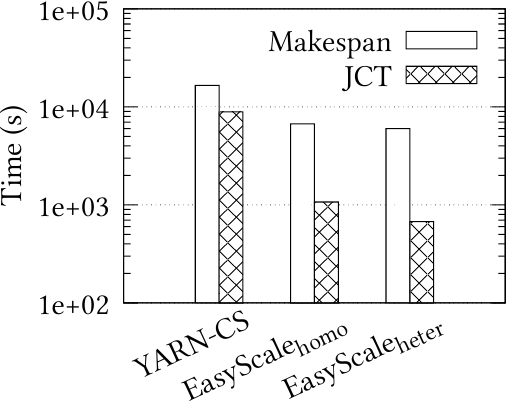} 
        \caption{Comparison of YARN-CS, \oursys{}\textsubscript{homo}, and \oursys{}\textsubscript{heter}.}
        \label{fig:trace_exp_all} 
    \end{minipage}
    \begin{minipage}[t]{0.32\linewidth}
        \centering 
        \includegraphics[width=0.85\linewidth]{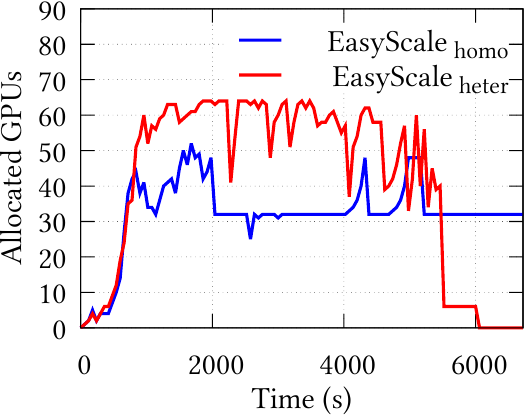} 
        \caption{Allocated GPUs of \oursys{}\textsubscript{homo} and \oursys{}\textsubscript{heter}.}
        \label{fig:trace_alloc} 
    \end{minipage}
    \begin{minipage}[t]{0.32\linewidth}
        \centering 
        \includegraphics[width=0.85\linewidth]{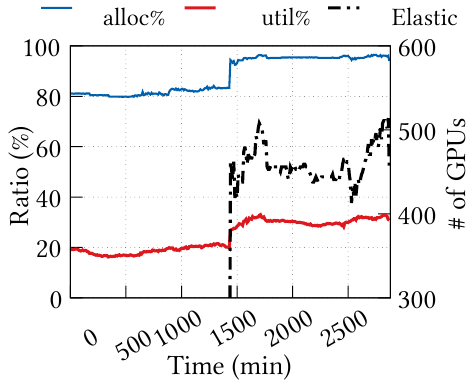} 
        \caption{One-day statistic on a production cluster of \ourcompany{}.} 
        \label{fig:clusterexp} 
    \end{minipage}%
\end{figure*}

\textbf{Workloads --}
The training jobs are configured as the DL workloads in Table~\ref{tab:trace_workloads}.
The job arrival time of the trace is configured according to Microsoft~\cite{gandiva2018}, 
and the job runtime distribution is down-sampled from our production training jobs.
All jobs are considered to be submitted to the same tenant with the same priority.

\textbf{Settings --}
We compare \oursys{} to Apache YARN's capacity scheduler (YARN-CS), a production GPU cluster scheduler used in Microsoft Philly~\cite{jeon2019analysis}. YARN-CS enforces FIFO mode for processing jobs to ensure inter-job fairness. In the experiment, all jobs use gang-scheduling to allocate GPU resources, and the minimal number of GPUs was set to 0 in \oursys{}. YARN-CS allocates the same type of GPUs (\eg{}, all V100 GPUs) for a job based on its requirement. And \oursys{} has two configurations: \textit{1)} \oursys{}\textsubscript{homo}, where a job can only use homogeneous GPUs by constraining scheduling plans of intra-job scheduler to homogeneous GPUs; and \textit{2)} \oursys{}\textsubscript{heter}, where a job can use heterogeneous GPUs.

\textbf{Results --}
Figure~\ref{fig:trace_exp_all} shows the average job completion time (JCT) and the makespan of different schedulers when scheduling the same job trace. \oursys{}\textsubscript{homo} and \oursys{}\textsubscript{heter} is compared to the capacity scheduler. 
The results show that \oursys{}\textsubscript{homo} improves average JCT by $8.3\times$ and makespan by $2.5\times$, while \oursys{}\textsubscript{heter} improves by $13.2\times$ and $2.8\times$.
Enabling elasticity eliminates the gang-scheduling requirement for training jobs, which significantly enhances performance through incremental utilization of idle GPUs. This results in a speedup as shown in \oursys{}\textsubscript{homo}. With the ability to utilize heterogeneous GPU resources, \oursys{}\textsubscript{heter} can utilize more available GPU resources. 
During execution, the allocated GPUs of \oursys{}\textsubscript{heter} are generally higher than those of \oursys{}\textsubscript{homo}. 
By ensuring consistent accuracy using heterogeneous GPU resources for elastic training, \oursys{} jobs can further utilize available GPUs of other types to achieve better throughput.

\subsection{Cluster Experiment}
\label{subsec:cluster_exp}

We have deployed \oursys{} in a shared GPU cluster with more than 3,000 GPUs.
This production cluster used to dedicate for online GPU serving or development (\eg, Jupyter Notebook).
Similar to Borg~\cite{borg}, which classifies jobs as production jobs (\ie, high-priority) and non-production batch jobs, we treat inference serving jobs as production jobs with guaranteed quota and treat \oursys{} jobs as non-production jobs to utilize the idle GPUs.

To illustrate the cluster efficiency improvement from \oursys{}, one-day statistic is collected in Dec. 2021, right after \oursys{} is fully deployed in this cluster.
As shown in Figure~\ref{fig:clusterexp}, the first 1,440 minutes (day-1) indicate the statistic collected before the deployment of \oursys{}, while the last 1,440 minutes (day-2) show how \oursys{} utilize the idle GPUs. On day-2, EasyScale jobs are submitted to this cluster according to the business patterns in real-world applications. These jobs contain different training workloads (CV/NLP) with different hyper-parameter settings. On average, \oursys{} improves the GPU allocation ratio by 17.1\% and improves the GPU SM utilization by 62.1\%. 
During the one-day statistic, the elastic \oursys{} jobs use 459 temporally idle GPUs on average that can quickly scale in to release GPUs for high-priority online serving jobs in seconds.
After the leaving of those inference jobs, \oursys{} jobs full up the idle GPUs within 5 minutes. 
Our cluster statistic records a total number of 362 preemptions on that day and no \oursys{} job fails.


\section{Related Work}
\label{sec:related}

\textbf{Determinism and reproducibility --} Determinism, reproducibility, and ablation study are important for DL researches~\cite{qian2021my, pham2020problems, nagarajan2018deterministic}.
DL frameworks~\cite{pytorchreproducibility} and NVIDIA~\cite{frameworkdeterminism} have studied deterministic model training using single GPU.
However, it is a hard problem due to the floating-point dominating execution, complicated software stack, and hardware optimized implementations. 
The design of \oursys{} derives from the understanding of non-determinism, considers bitwise identical in every step of model training, and extends the determinism to elastic training over heterogeneous GPUs.
In Ampere GPU (\eg, A100), cuBLAS supports only internal heuristics approach without public interface to select low-level kernel implementation, which can hardly produce accuracy-consistent results compared to that of using previous generation of GPUs~\cite{cublasa100}.

\textbf{Elastic deep learning --} TorchElastic~\cite{torchelastic}, ElasticDL~\cite{elasticdl}, and Horovod Elastic~\cite{horovodelastic} support elastic training and fault tolerance, however, they introduce non-determinism in model accuracy.
KungFu~\cite{kungfu2020} and Pollux~\cite{pollux2021} support adjusting training algorithms, including both adaptive batch sizes and learning rates, allowing both customized and build-in adaptation policies for efficient scaling.
VirtualFlow~\cite{virtualflow2021} and Varuna~\cite{varuna21} leverage the gradient accumulation approach to achieve elasticity. 
Those works cannot guarantee the trained model with consistent accuracy among different runs. 
As our parallel works, AutoPS~\cite{ps22}, Singularity~\cite{shukla2022singularity}, and Pathways\cite{barham2022pathways} also explore elastic training in different ways, including the model aggregations in parameter server architecture, {CUDA} calls analytics, and heterogeneous interconnects.
\oursys{} utilizes the DL characteristics to achieve efficient and accuracy-consistent elastic training.
\oursys{} currently focuses on data parallel, however, new parallel strategies are proposed for large model training~\cite{narayanan2019pipedream, fan2021dapple, jia2020whale, rajbhandari2020zero}, and we consider supporting them as future works.

\textbf{Cluster scheduling --} Resource management for DL jobs has been studied to improve utilization~\cite{gandiva2018,antman2020,weng2022mlaas} and fairness~\cite{gavel2020}. 
SLAQ~\cite{slaq17} and Pollux~\cite{pollux2021} prioritize resources by considering model convergence.
To improve cluster utilization, ONES~\cite{ebs2021} tunes batch size of training jobs.
Optimus~\cite{optimus} and EDL~\cite{edl2022} adjust the number of parameter-servers and workers.
PipeSwitch~\cite{pipeswitch2020} overlaps computation with layered model loading.
Retiarii~\cite{retiarii2020} dynamically allocates resources among AutoML jobs and applies cross-job optimization.
Gandiva~\cite{gandiva2018} and AntMan~\cite{antman2020} utilize the unique DL characteristic to optimize scheduling at mini-batch boundaries.


\section{Conclusion}
\label{sec:conclusion}


Through EasyScale, we demonstrate the success of decoupling DL training process from underlying resource allocation for achieving accuracy-consistent model training under elasticity. Specifically, EasyScale presents several innovations to address non-determinism during elastic training by
\textit{1)} introducing the EST abstraction to preserve the training behaviors over elasticity and heterogeneity, 
\textit{2)} sourcing the non-deterministic behaviors scattered in the DLT software stack and solving them, 
and \textit{3)} developing intra-job and inter-job schedulers utilizing the heterogeneous GPU cluster. 
Going forward, we hope EasyScale can draw attention to the deterministic computation of DL, and we should not always trade determinism for performance when designing DL systems.

\begin{acks}
\label{sec:ack}
We would like to thank the anonymous reviewers for their valuable comments and suggestions. We would also like to thank Ziheng Wu, Zhen Zheng for the valuable comments on the early version of this paper. This work is supported by National Natural Science Foundation of China (No. 62322201, 62072018 and U22A2028), and the Fundamental Research Funds for the Central Universities. Hailong Yang is the corresponding author.

\end{acks}


\bibliographystyle{ACM-Reference-Format}
\bibliography{reference}

%
%
%
%
%
%
%
%
%
%
\end{document}